\documentclass{article}
\usepackage[utf8]{inputenc}
\usepackage{authblk}
\usepackage{setspace}
\usepackage[margin=1.25in]{geometry}
\usepackage{graphicx}
\graphicspath{ {./figures/} }
\usepackage{subcaption}
\usepackage{amsmath}
\usepackage{lineno}
\usepackage[dvipsnames]{xcolor}
\usepackage{hyperref}
\hypersetup{
    colorlinks=true,
    linkcolor=cyan,
    filecolor=cyan,      
    urlcolor=cyan,
    citecolor=cyan,
    pdfpagemode=FullScreen,
    }

\title{\Large \textbf{A tripartite entanglement in de Sitter spacetime}}

\author[1]{Sang-Eon Bak}
\author[2]{Paul M. Alsing}
\author[3]{Warner A. Miller}
\author[3]{Shahabeddin M. Aslmarand}
\author[3,4*]{Doyeol Ahn}

\affil[1]{\textit{Department of Physics, Arizona State University, Tempe, AZ 85287, USA}}
\affil[2]{\textit{Air Force Research Laboratory, Information Directorate, Rome NY 13441, USA}}
\affil[3]{\textit{Physics Department, Charles E Schmidt College of Science, Florida Atlantic University, Boca Raton, FL 33431-0991, USA}}
\affil[4]{\textit{Department of Electrical and Computer Engineering, University of Seoul, 163 Seoulsiripdae-ro, Dongdaemun-gu, Seoul 02504, Korea}}
\affil[*]{Corresponding author: dahn@uos.ac.kr}

\date{}

\begin{document}

\maketitle

\begin{abstract}
We investigate the quantum correlation for tripartite entangled states in de Sitter space. First, we adopt the noisy quantum channel model. In this model, the expansion effect is represented by an operator sum representation with its corresponding Kraus operator. This map is shown to be trace-preserving and completely positive. Second, we analyze the quantum correlation by using the channel-state correspondence. For a large expansion rate, the tripartite mutual information has a large negative value, which corresponds to a small magnitude of bipartite mutual information. We relate this result with the challenge of recovering information from local measurements.
\end{abstract}


\section{Introduction}

In recent years, there has been growing interest in the connections between quantum information, gravity, and cosmology. Quantum information theory \cite{nielsen2002quantum,PhysRevA.54.2614} has been applied to the study of black hole information loss problems \cite{Alsing:2003es,Fuentes-Schuller:2004iaz,Alsing:2006cj,DoyeolAhn:2017jld,DoyeolAhn:2018tyo}. It was shown that black hole horizons can be modeled as noisy quantum channels \cite{DoyeolAhn:2017jld,DoyeolAhn:2018tyo}. Researchers are exploring concepts of entanglement and their correlation in the context of de Sitter space and its horizon \cite{Ball:2005xa,Fuentes:2010dt,Nambu:2011ae,Alsing:2005dno,Kanno:2014ifa}. The impact of quantum entanglement on cosmological quantum fluctuation has also been investigated \cite{Kanno:2014ifa}. Other useful references are in order \cite{Maldacena:2012xp,Choudhury:2016cso,Choudhury:2016pfr,Choudhury:2017bou,Choudhury:2017qyl,Choudhury:2018ppd,Bohra2021,Akhtar:2019qdn,Banerjee:2020ljo,Choudhury:2022mch,Kanno:2014lma,Kanno:2014bma,Kanno:2015ewa,Kanno:2016gas,Kanno:2016qcc,Kanno:2017dci,Albrecht:2018prr,Kanno:2021gpt,Kanno:2022kve,Torres-Arenas:2018vei,Torres-Arenas_2019,Quantum_2019,Dong_2019,Dong:2019iqt,Dong:2022vfw}.

However, there is still a lack of understanding of the properties of the tripartite entanglement distributed in the de Sitter space. In this paper, we explore the quantum correlation for tripartite entangled states, concentrating on massless scalar fields interacting with the de Sitter space. We adopt a quantum information theoretic perspective by determining the behavior of quantum entanglement across the de Sitter horizon. First, we model a de Sitter horizon as a noisy quantum channel by employing the technique in \cite{DoyeolAhn:2017jld, DoyeolAhn:2018tyo}. Based on this model, we computed the measure of entanglement, which is known as fidelity. Second, we study the mutual information of the tripartite states and the information exchange with an observer beyond the horizon by using the channel-state correspondence \cite{Hosur:2015ylk}. We analyze the entangled states with various information-theoretic quantities: bipartite mutual information, tripartite mutual information, and Renyi-2 version of mutual information. We then relate our results with the scrambling effect, which indicates the difficulties of recovering information from local measurements on the subsystems.

The paper is organized as follows. In section \ref{section_2}, we consider the quantization of the massless scalar field in de Sitter spacetime, adopting the formalism of Gibbons and Hawking \cite{Gibbons:1977mu} and Lohiya and Panchapakesan \cite{Lohiya:1978}. Based on this formalism, we introduce our setup: the entangled tripartite quantum states across the horizon. Following that, in section \ref{section_3}, we will study the entangled state in two different ways: (i) using a noisy quantum channel model, and (ii) using channel-state correspondence. In section \ref{section_4}, we provide the conclusion. We relegate a review of the channel-state correspondence to Appendix \ref{Appendix_A}.

\section{Setup}\label{section_2}
\subsection{Massless scalar fields in static de Sitter spacetime}
We will consider the massless scalar field in (3+1) dimensional static de Sitter spacetime. The static de Sitter metric is defined by
\begin{equation}\label{eq_de Sitter metric}
d s^{2}=-\left(1-\frac{r^{2}}{a^{2}}\right) d t^{2}+\left(1-\frac{r^{2}}{a^{2}}\right)^{-1} d r^{2}+r^{2}\left(d \theta^{2}+\sin ^{2} \theta d \phi^{2}\right)\,,
\end{equation}
where $a$ is the location of the horizon in de Sitter spacetime, and $a=\sqrt{3 / \Lambda}$ where $\Lambda$ is the cosmological constant. In this coordinate, the light signal which starts at the position $r>a$ cannot reach an observer at $r=0$.

Then, we can define the tortoise coordinate.
\begin{equation}
r^{*}=\frac{a}{2} \ln \left(\frac{a+r}{a-r}\right)\,.
\end{equation}
With the tortoise coordinate, the static de Sitter metric can be rewritten as
\begin{equation}
d s^{2}=-\left(1-\frac{r^{2}}{a^{2}}\right) d \tilde{u} d \tilde{v}+r^{2}\left(d \theta^{2}+\sin ^{2} \theta d \phi^{2}\right)\,,
\end{equation}
where the light cone coordinates $\tilde{u}=t-r^{*}, \tilde{v}=t+r^{*}$ are applied. Then, by using the coordinates $u=$ $-2 a e^{\tilde{u} / a}, v=2 a e^{-\tilde{v} / a}$, the metric in Kruskal coordinates can be given as the following form.
\begin{equation}
d s^{2}=\frac{1}{4}\left(1+\frac{r^{2}}{a^{2}}\right) d u d v+r^{2}\left(d \theta^{2}+\sin ^{2} \theta d \phi^{2}\right)\,.
\end{equation}

Then, we can construct a Kruskal diagram in Figure 1 by using the following coordinates.
\begin{equation}
T=\frac{u+v}{2}=-2 a e^{\frac{r^{*}}{a}} \sinh \frac{t}{a}, R=\frac{u-v}{2}=2 a e^{\frac{r^{*}}{a}} \cosh \frac{t}{a}\,.
\end{equation}

\begin{figure}
\begin{center}
\begin{tabular}{cc}
\setlength{\unitlength}{1cm}
\hspace{0.1cm}
\includegraphics[width=6.5cm]{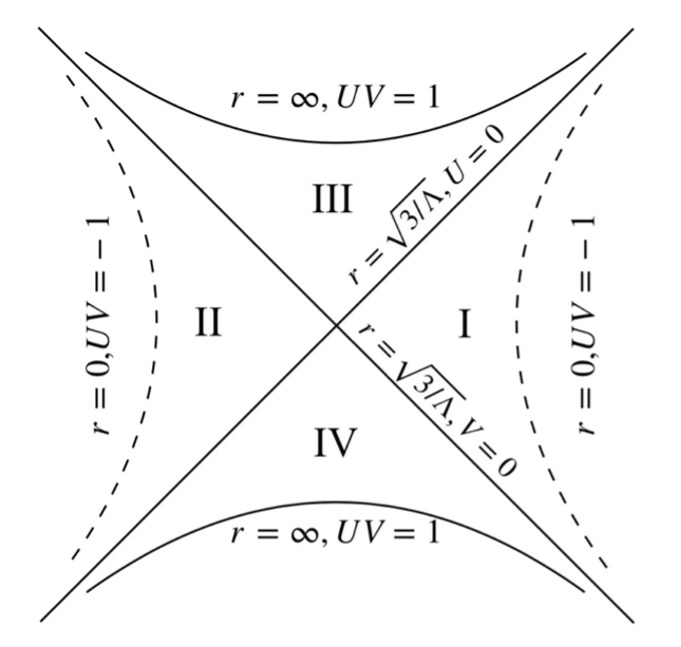}
\qquad &
\qquad
\includegraphics[width=6.5cm]{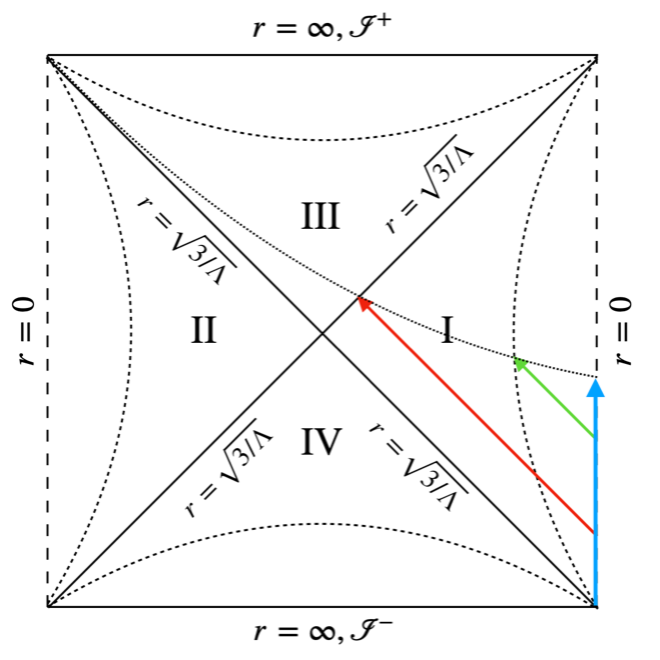}
\qquad
\end{tabular}
\end{center}
\caption{The Kruskal diagram (Left) and the Penrose diagram (Right) of the de Sitter spacetime. The red arrow indicates Alice’s trajectory, and the blue arrow indicates Bob’s trajectory. The green arrow corresponds to Charlie’s trajectory.}
    \label{fig-Penrose}
\end{figure}

Now, we consider the massless scalar field in de Sitter spacetime. The field quantization in de Sitter spacetime is constructed by Gibbons and Hawking \cite{Gibbons:1977mu}, and Lohiya and Panchapakesan \cite{Lohiya:1978}. Here, we follow the method discussed in \cite{Lohiya:1978}. To follow a similar quantization procedure with the Schwarzschild case, we consider the conformally coupled scalar field. The conformally coupled scalar field equation is given by
\begin{equation}\label{eq_field eq}
(-g)^{-\frac{1}{2}} \frac{\partial}{\partial x^{\mu}}\left[g^{\mu \nu}(-g)^{\frac{1}{2}} \frac{\partial}{\partial x^{v}}\right] \phi=\frac{1}{6} R \phi\,.
\end{equation}
First, we consider the equation in the light cone coordinate. Here, the separation of variables in the mode equation is considered. The de Sitter metric in eq. (\ref{eq_de Sitter metric}) is static and spherically symmetric. Using these facts, radial and angular, and time dependencies in the solution of (\ref{eq_field eq}) can be separated. Referring to \cite{Lohiya:1978}, the light cone mode solution is written in the following form.
\begin{equation}
\phi_{\omega l m}(r, \theta, \phi)=\frac{f_{\omega l}(r)}{r} Y_{l m}(\theta, \phi) e^{-i \omega t}\,.
\end{equation}
By some mathematical manipulation, we can get the radial equation with the following form where $z=$ $a / r$.
\begin{equation}
\left(1-z^{2}\right) \frac{d}{d z}\left(\left(1-z^{2}\right) \frac{d f_{\omega l}}{d z}\right)+\left[a^{2} \omega^{2}+\left(l(l+1)\left(1-z^{2}\right)\right)\right] f_{\omega l}=0\,.
\end{equation}
Since the radial solutions behave as $e^{ \pm i \omega r^{*}}$ in the $r \rightarrow a$ limit, the light cone mode solution can be written as the incoming and outgoing waves at the horizon.

\begin{equation}
\frac{Y_{l m}}{r} e^{-i \omega \tilde{u}}, \quad \frac{Y_{l m}}{r} e^{-i \omega \tilde{v}}\,.
\end{equation}
We proceed by utilizing the mode expansion approach presented in Birrell and Davies \cite{Birrell:1982ix} and Unruh \cite{Unruh:1976db} to solve the conformally coupled massless Klein-Gordon equation within the context of static de Sitter spacetime. In their approach, the light cone modes for regions I and II are non-analytic at $u=v=0$. Consequently, these light cone modes exhibit different vacuum states compared to the Kruskal mode. Nevertheless, similar to the Schwarzschild case, we can represent the positive frequency Kruskal modes, which are analytic at $u=v=0$, as a combination of incoming and outgoing waves of the light cone modes for regions I and II.

The positive frequency mode for the Kruskal coordinate is given as
\begin{equation}
\frac{1}{\sqrt{2 \sinh (\pi a \omega)}}\left(e^{\pi a \omega / 2} u_{\omega l m}^{I}+e^{-\pi a \omega / 2} u_{\omega l m}^{I I}\right)\,,
\end{equation}
where $u_{\omega l m}^{I}$ is the light cone mode vanishing for $u<0$, and $u_{\omega l m}^{I I}$ is the light cone mode vanishing for $u>0$. This approach which is noted in \cite{Unruh:1976db} is firstly suggested by Unruh for deriving the relation between two different vacuums for each mode in Rindler spacetime consideration. From this analysis, the Bogoliubov transformation for the relation between the vacuum state for the Kruskal mode and the vacuum state for the light cone mode can be derived as the following formula.
\begin{equation}
\begin{aligned}
& d_{\omega l m}=\cosh \gamma b_{\omega l m}^{I I}-\sinh \gamma b_{\omega l m}^{\dagger}\,, \\
& d_{\omega l m}^{\dagger}=\cosh \gamma b_{\omega l m}^{\dagger}-\sinh \gamma b_{\omega l m}^{I I}\,,
\end{aligned}
\end{equation}
where $\tanh \gamma=e^{-\pi a \omega}$. Here, $d_{\omega l m}, d_{\omega l m}^{\dagger}$ are the annihilation and creation operator for the Kruskal mode, $b_{\omega l m}^{I}, b^{\dagger}{ }_{\omega l m}^{I}$ is the annihilation and creation operator for the light cone mode in region I. The creation and annihilation operators for the light cone mode in region II are defined by changing its upper indices. These annihilation operators and creation operators generate their corresponding vacuum states respectively. The vacuum state of the Kruskal mode is defined by
\begin{equation}
d_{\omega l m}|0\rangle_{K}=0\,,
\end{equation}
where $|0\rangle_{K}$ is the vacuum state for Kruskal mode. The vacuum state of light cone mode is defined by
\begin{equation}
b_{\omega l m}^{I}|0\rangle_{I}=b_{\omega l m}^{I I}|0\rangle_{I I}=0\,,
\end{equation}
where $|0\rangle_{I},|0\rangle_{I I}$ are the vacuum states for light cone mode in regions I and II respectively. These yield the corresponding Fock spaces. By using the Bogoliubov transformation, the vacuum state of the acceleration frame can be described as a two-mode squeezed state for the observer staying at the origin. We adhere to the mathematical techniques presented in \cite{Alsing:2003es,Fuentes-Schuller:2004iaz}.

\begin{equation}\label{eq_vacuum}
|0\rangle_{K}=\frac{1}{\cosh \gamma} \sum_{n=0}^{\infty} \tanh ^{n} \gamma|n\rangle_{I}|n\rangle_{I I}\,.
\end{equation}
This finding is interpreted as the phenomenon in which an observer in the region I, with the light cone mode and its associated vacuum state, can detect particle creation close to the horizon. Similarly, we derive the first excited state in the Fock space for the Kruskal mode as the following result \cite{Alsing:2003es,Fuentes-Schuller:2004iaz}.
\begin{equation}\label{eq_excited}
|1\rangle_{K}=\frac{1}{\cosh ^{2} \gamma} \sum_{n=0}^{\infty} \tanh ^{n} \gamma \sqrt{n+1}|n+1\rangle_{I}|n\rangle_{I}\,.
\end{equation}
By using these results, we analyze entanglement in de Sitter spacetime.

\subsection{Entangled states in static de Sitter spacetime}

In our model, there are three observers who initially share their quantum states (See, Figure 1-(b)). First, Bob remains stationary at the origin and observes radiation at the horizon. Second, Alice moves away from Bob at a speed equal to the expansion rate of de Sitter spacetime, eventually passing through the horizon from Bob's perspective. Third, Charlie also moves away from Bob at the same speed as Alice, but at a later time, and does not cross the horizon according to Bob's observation. We choose this trajectory for Alice because we want to investigate the entanglement across the horizon. Next, let's consider the states that the three observers share. Bob's state corresponds to the quantum state generated by the creation operator for light cone mode. Alice and Charlie share a state that corresponds to the quantum state generated by the creation operator for the Kruskal mode. Therefore, Alice and Charlie cannot detect the effect of radiation and stay in our expanding universe from Bob's point of view.

Moreover, we treat Charlie as the reference state. The combined state of Alice and Bob, after Alice crosses the horizon, is not a pure state because Alice's system must be traced out for observations within the universe. In this situation, it is reasonable to consider a reference system that remains in our universe. With the reference system in place, Bob can physically verify the success of his decoding of Alice's state. This is because we can interpret that Bob possesses complete information about Alice's state when Bob's state becomes a purification of Charlie's state. In relation to the decoding process, we can calculate the correlation between Alice and Bob, as well as the scrambling effect in the channel. This can be studied by introducing GHZ and W states in the following. Initially, we consider two kinds of tripartite entangled states, which are written in the following forms.
\begin{align}
&|\psi\rangle^{(G H Z)}  =\frac{1}{\sqrt{2}}(|000\rangle+|111\rangle)\label{eq_GHZ state}\,,\\
&|\psi\rangle^{(W)}  =\frac{1}{\sqrt{3}}(|100\rangle+|010\rangle+|001\rangle)\label{eq_W state}\,,
\end{align}
where we denote the notation as $|000\rangle=\left|0_{A}\right\rangle_{K}\left|0_{B}\right\rangle_{K}\left|0_{C}\right\rangle_{K}$ and the states $\left|0_{A}\right\rangle_{K},\left|0_{B}\right\rangle_{K},\left|0_{C}\right\rangle_{K}$ represent the Kruskal vacuum states that can only be detected by Alice, Bob, and Charlie respectively. There are two well-known types of entangled states: one is referred to as the Greenberger-Horne-Zeilinger (GHZ) state, and the other is known as the W state. As a result of the universe's expansion, Bob's vacuum state and first excited states in equations (\ref{eq_GHZ state}) and (\ref{eq_W state}) evolve into a two-mode squeezed state in Fock space for light cone modes (\ref{eq_vacuum}) and (\ref{eq_excited}). The resultant states are given by
\begin{equation}\label{eq_ent 1}
    |\psi'\rangle^{(GHZ)}=\frac{1}{\sqrt{2}}\sum_n\frac{\tanh^n{\gamma}}{\cosh^2{\gamma}}\left(\cosh{\gamma}|0\,0\,n\,n\rangle+\sqrt{n+1}|1\,1\,n+1\,n\rangle\right)
\end{equation}
\begin{equation}\label{eq_ent 2}
    |\psi'\rangle^{(W)}=\frac{1}{\sqrt{3}}\sum_n\frac{\tan ^n{\gamma}}{\cosh^2{\gamma}}\left(\cosh{\gamma}|1\,0\,n\,n\rangle+\cosh{\gamma}|0\,1\,n\,n\rangle+\sqrt{n+1}|0\,0\,n+1\,n \right)
\end{equation}
where $|i\,j\,n\,m\rangle=\left|i_{C}\right\rangle_{K}\left|j_{A}\right\rangle_{K}\left|n_{B}\right\rangle_{I}\left|m_{B}\right\rangle_{II}$ for $i,j=0,1$, $n,m=0,1,2,\cdots$ are the quantum states for Charlie, Alice, and Bob's states in regions I and II respectively. The entangled states of our interest are the tripartite state since we will trace out Bob's mode II.

\section{Methods and results}\label{section_3}
\subsection{Noisy quantum channel}\label{section 3.1}
In this section, we will analyze the entangled states (\ref{eq_ent 1}) and (\ref{eq_ent 2}) by using the noisy quantum channel model. To do this, we establish the connection between the initial state and the final state. 

When the de Sitter universe's expansion begins, Bob experiences particle creation, and the three observers initiate their respective movements. This evolved state is treated as the final state. Utilizing the final quantum state, the outcome can be represented by the density matrix of a mixed state. For describing the evolution of the entanglement, we consider the density matrix $\rho=|\psi\rangle\langle\psi|$. We trace out Charlie's mode because we are interested in the quantum correlation between Bob and Alice. Then, by using equations (\ref{eq_GHZ state}), (\ref{eq_W state}) density matrices of the initial states are given in the following form.
\begin{gather}
\rho_{A B}^{(G H Z)}=\operatorname{Tr}_{C}(|\psi\rangle\langle\psi|)=\frac{1}{2}(|00\rangle\langle 00|+| 11\rangle\langle 11|)\,, \\
\rho_{A B}^{(W)}=\operatorname{Tr}_{C}(|\psi\rangle\langle\psi|)=\frac{1}{3}(|10\rangle\langle 10|+| 10\rangle\langle 01|+| 01\rangle\langle 10|+| 01\rangle\langle 01|+| 00\rangle\langle 00|)\,.
\end{gather}
Since there are two causally disconnected regions in de Sitter spacetime and Bob can exist only in region I, we should trace out one of Bob's modes in region II in the density matrices of the final state. Subsequently, the final state we established earlier is as follows:
\begin{equation}\label{eq_Thermal GHZ}
\rho_{AB}^{\prime(GHZ)}=\frac{1}{2 \cosh ^{2} \gamma} \sum_{n=0}^{\infty} \tanh ^{2 n} \gamma\left[|0 n\rangle\langle 0 n|+\frac{n+1}{\cosh ^{2} \gamma}| 1 n+1\rangle\langle 1 n+1|\right]\,,
\end{equation}
\begin{equation}\label{eq_Thermal W}
\begin{aligned}
 \rho_{A B}^{\prime(W)}=&\frac{1}{3 \cosh ^{2} \gamma} \sum_{n}^{\infty} \tanh ^{2 n} \gamma\bigg[|1 n\rangle\langle 1 n|+| 0 n\rangle\langle 0 n|+\frac{(n+1)}{\cosh ^{2} \gamma}| 0 n+1\rangle\langle 0 n+1| \\
& +\frac{\sqrt{n+1}}{\cosh \gamma}|1 n\rangle\langle 0 n+1|+\frac{\sqrt{n+1}}{\cosh \gamma}| 0 n+1\rangle\langle 1 n|\bigg]\,.
\end{aligned}
\end{equation}

We establish the relation between the initial state and the final state. The relation implies that the expansion effect can be represented as an operator sum representation.
\begin{equation}
\varepsilon\left(\rho_{A B}\right)=\sum A_{n} \rho_{A B} A_{n}^{\dagger}=\rho_{A B}^{\prime}\,.
\end{equation}
The density matrices $\rho_{A B}$ and $\rho_{A B}^{\prime}$ describe the initial and final state of our model. Here, we can derive the Kraus operator as the following form for both the W state case and the GHZ state case.
\begin{equation}
A_{n}=\frac{1}{\sqrt{n !}} \frac{\tanh ^{n} \gamma}{\cosh \gamma}\left(b_{I}^{+}\right)^{n} \otimes \frac{1}{(\cosh \gamma)^{b_{I}^{+} b_{I}}}\,.
\end{equation}
We confirm that this map is trace-preserving and completely positive. For this reason, we can treat these maps as a noisy quantum channel \cite{PhysRevA.54.2614}.

\paragraph{Fidelity} In this setup, we calculate the fidelity of the channel. The Kraus operator is used as an easier approach to calculate the fidelity of the channel, which can be represented by the formula $F_{e}=$ $\sum_{n}\left(tr\left(\rho_{A B} A_{n}\right) tr\left(\rho_{A B} A_{n}^{\dagger}\right)\right)$

\begin{align}
F_{e}^{(G H Z)} & =\frac{1}{4 \cosh ^{2} \gamma}\left(1+\frac{1}{\cosh ^{2} \gamma}\right)^{2}\,, \\
F_{e}^{(W)} & =\frac{1}{9 \cosh ^{2} \gamma}\left(2+\frac{1}{\cosh ^{2} \gamma}\right)^{2}\,.
\end{align}
The fidelity $F_{e}={ }_{A B}\left\langle\psi\left|\rho_{A B}^{\prime}\right| \psi\right\rangle_{A B}$ represents the preservation of entanglement when they transformed from the initial state to the final state. In Figure 2, we show that the fidelities of the channel for the GHZ state and the W state cases decrease when the expansion rate grows. This effect can be interpreted as entanglement degradation. When we consider the non-zero expansion rate, the fidelity has a value of less than 1. Since the fidelity is related to the entanglement, the result shows that entanglement is degraded in the presence of non-zero expansion. Our result agrees with the fidelity obtained in non-inertial frames \cite{Alsing:2003es,DoyeolAhn:2017jld}.

\begin{figure}
    \centering
    \includegraphics[width=0.6\textwidth]{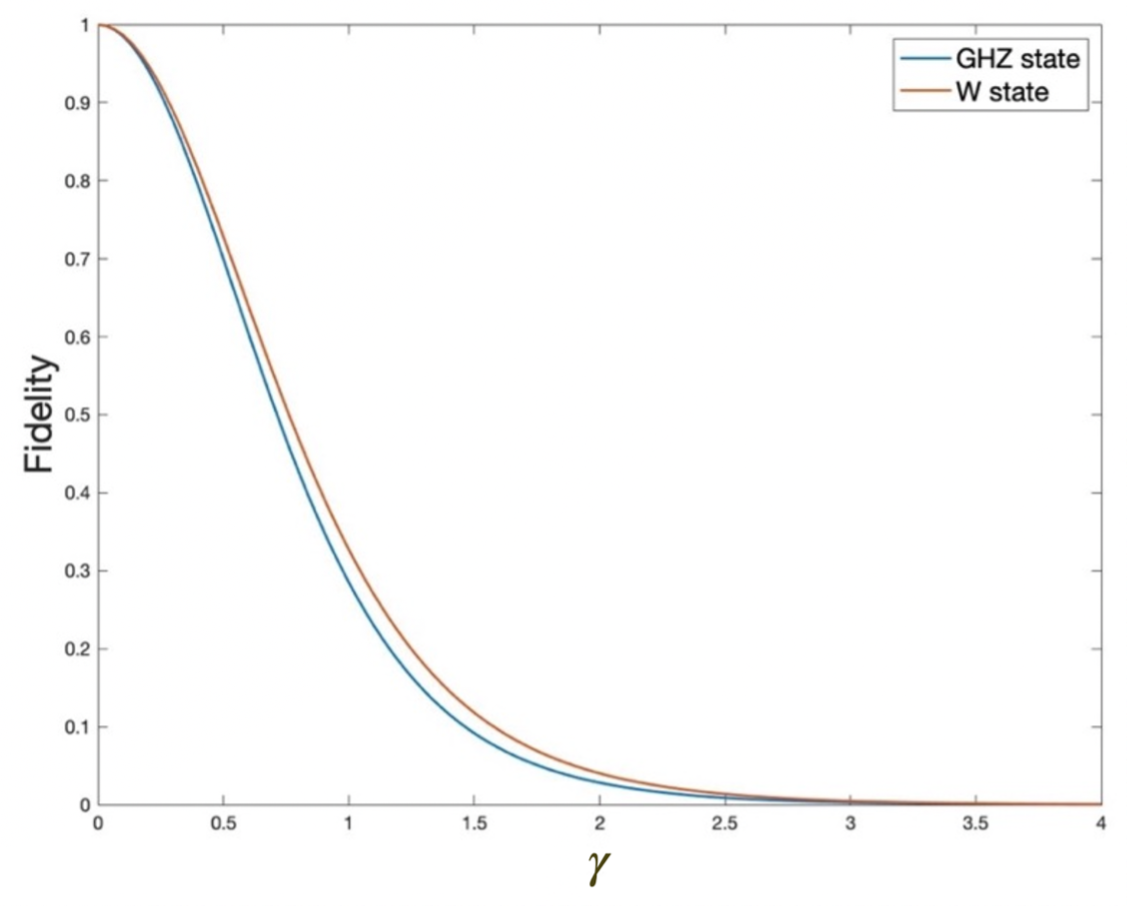}
    \caption{The red line indicates the fidelity of the thermalized state with its initial state as W state, and the blue line indicates the fidelity of the thermalized state with its initial state GHZ state. These numerical results of fidelity decrease as a function of expansion rate $\gamma$.}
    \label{fig-2}
\end{figure}

\subsection{Channel-state correspondence}\label{section 3.2}
In this section, independent of the approach in section \ref{section 3.1}, we will study the property of the entangled states (\ref{eq_ent 1}) and (\ref{eq_ent 2}) by using the channel-state correspondence. The channel-state correspondence is used to establish a connection between quantum channels and quantum states. A review of channel-state correspondence is included in Appendix \ref{Appendix_A}. In this section, we are interested in the following state 
\begin{equation}\label{eq_tripartite state}
\rho_{CAB}'=tr_{II}\left(|\psi'\rangle\langle\psi'|\right)
\end{equation}
for each GHZ and W state in (\ref{eq_ent 1}) and (\ref{eq_ent 2}) respectively.
\paragraph{Mutual information} For the density matrix of the tripartite state in (\ref{eq_tripartite state}), we estimate the amount of the correlation between two quantum states by calculating the mutual information, which is defined as $I(A:B)=S(\rho_A)+S(\rho_B)-S(\rho_{AB})$. The bipartite mutual information can be studied to indicate the amount of Alice's information that Bob can obtain with his measurement of his state. When Alice, Bob, and Charlie share the entangled state, the bipartite mutual information is given as
\begin{equation}
\begin{aligned}
& I^{(G H Z)}(A: B)=1+\frac{1}{2 \cosh ^{2} r} \sum_{n=0}^{\infty} \tanh ^{2 n} \gamma\left[\left(W_{n}-2\right) \log _{2}\left(W_{n}-2\right)\right. \\
&\left.-\left(W_{n}-1\right) \log _{2}\left(W_{n}-1\right)\right]\,,
\end{aligned}
\end{equation}
\begin{equation}
\begin{aligned}
& I^{(W)}(A: B)=\log _{2} 3-\frac{5}{3}-\frac{1}{3 \cosh ^{2} \gamma} \log _{2}\left(3 \cosh ^{2} \gamma\right)-\frac{1}{3} \log _{2}\left(\tanh ^{2} \gamma\right) \\
&- \frac{1}{6 \cosh ^{2} \gamma} \sum_{n=0}^{\infty} \tanh ^{2 n} \gamma\left[M_{n}^{+} \log _{2} M_{n}^{+}+M_{n}^{-} \log _{2} M_{n}^{-}-2 W_{n} \log _{2} W_{n}\right. \\
&\left.-\frac{2}{\cosh ^{2} \gamma} \log _{2}\left(\frac{\tanh ^{2 n} \gamma}{6 \cosh ^{2} \gamma}\right)\right]\,,
\end{aligned}
\end{equation}
where 
\begin{equation*}
M_{n}^{ \pm}(\gamma)=1+\tanh ^{2} \gamma+\frac{n+1}{\cosh ^{2} \gamma} \pm\left(1-\tanh ^{2} \gamma+\frac{1}{\cosh ^{2} \gamma}\right)^{\frac{1}{2}}\left(1+3 \tanh ^{2} \gamma+\frac{n+1}{\cosh ^{2} \gamma}\right)^{\frac{1}{2}}\,,
\end{equation*}
\begin{equation*}
W_{n}(\gamma)=\left(2+\frac{n}{\sinh ^{2} \gamma}\right)\,,
\end{equation*}
which we plot in Figure 3. When the expansion rate grows, the bipartite mutual information for the prepared state decreases. The decrease of the bipartite mutual information starts from 1 for the GHZ state case. On the other hand, for the W state case, the decrease of the bipartite mutual information starts from approximately 2. 
\begin{figure}
    \centering
    \includegraphics[width=0.6\textwidth]{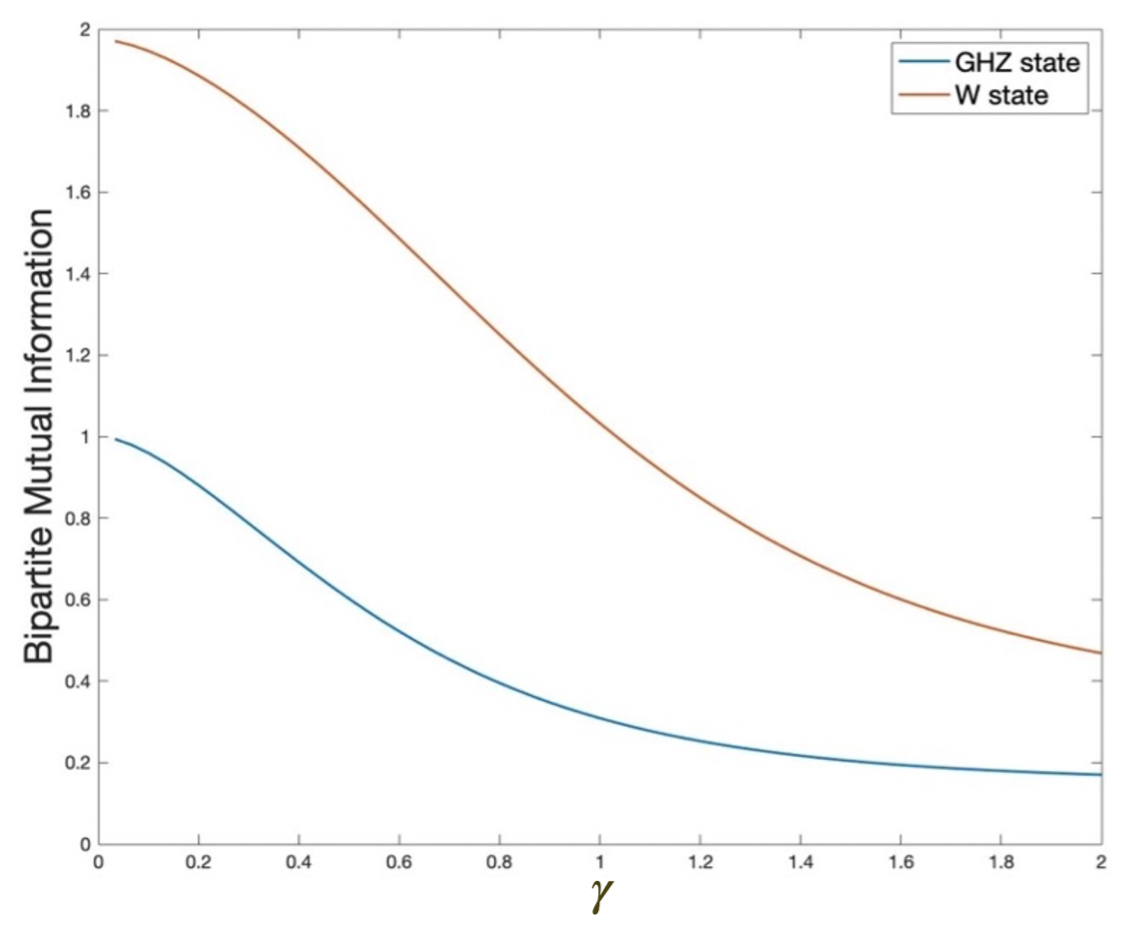}
    \caption{The red line indicates the bipartite mutual information of the thermalized state, which is initially the W state. The blue line indicates the bipartite mutual information of the thermalized state, which is initially the GHZ state. Bipartite mutual information for both states decreases as a function of expansion rate $\gamma$.}
    \label{fig-3}
\end{figure}

It is known that the GHZ state becomes unentangled after tracing out one part of the total state \cite{Dur:2000zz}. For the W state, the bipartite state resulting from tracing over the total state is also entangled. This implies that the W state is less susceptible to the loss of one quantum state than the GHZ state, even though the GHZ state exhibits a greater correlation across the entire tripartite system. For this reason, there is a distinction in the mutual information at the $\gamma \rightarrow 0$ limit between the cases of the GHZ state and the W state.

\paragraph{Tripartite mutual information} For the tripartite states in (\ref{eq_tripartite state}), we compute the tripartite mutual information, defined by $I(A: B: C)=I(A: B)+I(A: C)-I(A: B C)$,\footnote{We will write $B \cup C$ by $BC$ for convenience.} in conjunction with channel-state correspondence to assess the scrambling effect in de Sitter spacetime. After performing certain manipulations, the tripartite information for each state can be expressed as follows:
\begin{equation}\label{eq_TMI GHZ}
\begin{aligned}
&I^{(G H Z)}(A: B: C)=  1+\frac{1}{2 \cosh ^{2} \gamma} \sum_{n=0}^{\infty} \tanh ^{2 n} \gamma\left[\left(W_{n}-2\right) \log _{2}\left(W_{n}-2\right)\right. \\
& +\left(W^{\prime}{ }_{n}-2\right) \log _{2}\left(W^{\prime}{ }_{n}-2\right) \left.-\left(W_{n}-1\right) \log _{2}\left(W_{n}-1\right)-\left(W^{\prime}{ }_{n}-1\right) \log _{2}\left(W^{\prime}{ }_{n}-1\right)\right]\,,
\end{aligned}
\end{equation}
\begin{equation}\label{eq_TMI W}
\begin{aligned}
I^{(W)}(A: B: C)=\log _{2} 3- & \frac{8}{3}-\frac{2}{3 \cosh ^{2} \gamma} \log _{2}\left(3 \cosh ^{2} \gamma\right)-\frac{1}{3} \log _{2}\left(\tanh ^{2} \gamma\right) \\
& -\frac{1}{3 \cosh ^{2} \gamma} \sum_{n=0}^{\infty} \tanh ^{2 n} \gamma\bigg[M_{n}^{+} \log _{2} M_{n}^{+}+M_{n}^{-} \log _{2} M_{n}^{-}-W_{n} \log _{2} W_{n} \\
& -W^{\prime}{ }_{n} \log _{2} W^{\prime}{ }_{n}-\frac{2}{\cosh ^{2} \gamma} \log _{2} \frac{\tanh ^{2 n} \gamma}{6 \cosh ^{2} \gamma}\bigg] \,,
\end{aligned}
\end{equation}
where $W^{\prime}{ }_{n}(\gamma)=\left(2+\frac{n+1}{\cosh ^{2} \gamma}\right)$.
For the thermalized GHZ state, the tripartite mutual information decreases monotonically as the expansion rate grows, as depicted in Figure 4. This result is similar to the behavior of bipartite mutual information, which also exhibits a decreasing curve with an increasing expansion rate $\gamma$.
\begin{figure}
    \centering
    \includegraphics[width=0.6\textwidth]{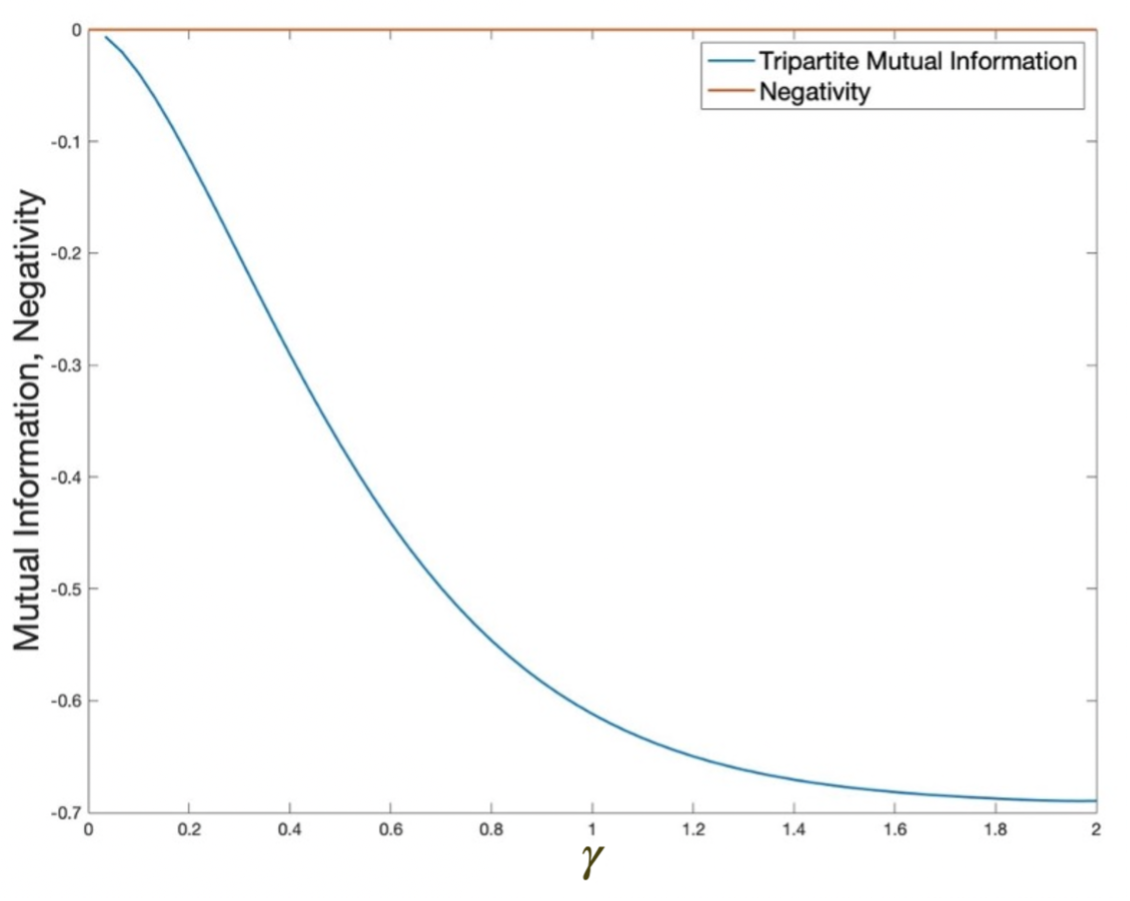}
    \caption{The blue line indicates the tripartite mutual information of the thermalized state with its initial state as GHZ state. The red line indicates the negativity of the state. The tripartite mutual information has always a negative value, but its absolute value increases as a function of expansion rate $\gamma$. There is no negative eigenvalue of $\rho_{AB}^{T_A}$ for all expansion rate $\gamma$.}
    \label{fig-4}
\end{figure}

\begin{figure}
    \centering
    \includegraphics[width=0.6\textwidth]{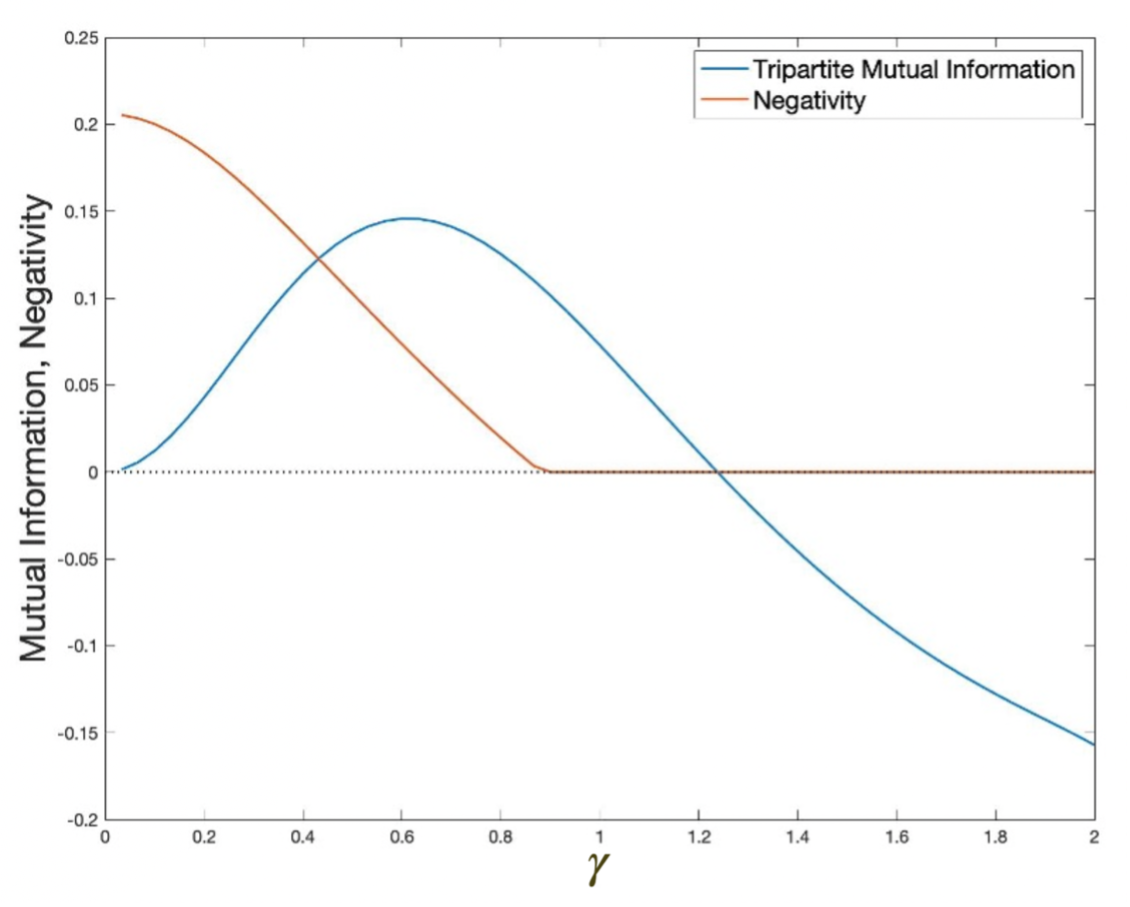}
    \caption{The blue line indicates the tripartite mutual information of the thermalized state with its initial state as W state. The red line indicates the negativity of the state. The tripartite mutual information has a region where it has a positive value, but its absolute value decreases when the expansion rate $\gamma$ is large. There are negative eigenvalues of $\rho_{AB}^{T_A}$ for some expansion rate $\gamma$.}
    \label{fig-5}
\end{figure}

For the W state case, however, there is non-trivial behavior that we can investigate in the plot. There is a positive value of tripartite mutual information in the thermalized W state on some range. We will analyze this non-trivial behavior by analyzing the relation with negativity \cite{Vidal:2002zz}, which is treated as a measure of entanglement. 

\paragraph{Negativity} For a density matrix $\rho_{A B}$, we can perceive that there is distillable entanglement when there is at least one negative eigenvalue of the matrix $\rho_{A B}^{T_{A}}$, which indicates the partial transpose of the density matrix. Otherwise, we cannot determine whether there is entanglement in the quantum state when there exist only positive eigenvalues. It is known as the partial transpose criterion.

For the GHZ state, only positive eigenvalues exist among eigenvalues of the matrix $\rho_{A B}^{T_{A}}$. On the other hand, eigenvalues of $\rho_{A B}^{T_{A}}$ for the W state case are given as the following, for $n=0,1,2, \ldots$
\begin{equation}
\lambda_n=\frac{\tanh ^{2 n} \gamma}{6 \cosh ^2 \gamma}\left[\left(1+\frac{n}{\sinh ^2 \gamma}+\tanh ^2 \gamma\right) \pm \sqrt{\left(1+\frac{n}{\sinh ^2 \gamma}+\tanh ^2 \gamma\right)^2-4 \tanh ^2 \gamma+\frac{4}{\cosh ^2 \gamma}}\right]\,.
\end{equation}
There exists a negative eigenvalue when $4 \tanh ^{2} \gamma+\frac{4}{\cosh ^{2} \gamma}$ is greater than 0 . We calculate numerically that $4 \tanh ^{2} \gamma+\frac{4}{\cosh ^{2} \gamma}$ is greater than 0 when the expansion rate $\gamma$ is less than $0.783\dots$ approximately. For the state with an expansion rate $\gamma$ less than $0.783\dots$, there remains a distillable entanglement in the state.

We plot the sum of the absolute value of negative eigenvalues $\sum_{n=0}^{\infty}\left|\lambda_{n}\right|$, which indicates the amount of negative eigenvalue with tripartite mutual information in Figures 4 and 5. In the plot, there are different properties between the ranges below and above $\gamma=0.783\dots$. It is similar to the behavior of the tripartite mutual information, which has a different sign between two distinct ranges.
However, in the region greater than $0.783\dots$, there is still a region with positive tripartite mutual information. In this region, we find that a bound entangled state might exist, as no distillable entangled states are present for these expansion rates. Since there are challenges in determining the existence of a bound entangled state \cite{Horodecki:2009zz}, we will not rigorously investigate the measure of bound entanglement. We leave this work as a subsequent study.\footnote{The entanglement measure of the tripartite system can be analyzed by comparing it to those of the reduced bipartite systems. In future work, we plan to study the entanglement measure using concurrence \cite{Hill:1997pfa,Rungta_2001}, an entanglement measure specifically for bipartite quantum states. Applying concurrence to each bipartite system in our model will help us gain a better understanding of the entanglement structure within the entire tripartite state.}


\paragraph{Renyi-2 version of tripartite mutual information} For the entangled states (\ref{eq_tripartite state}), we study the Renyi-2 version of tripartite mutual information. Then, we compare the behaviors of the Renyi-2 version of mutual information and tripartite mutual information. The Renyi-2 version tripartite mutual information is calculated by using the Renyi entropy $S^{(\alpha)}(\rho)=\frac{1}{1-\alpha} \log \operatorname{tr}\left(\rho^{\alpha}\right)$ with $\alpha=2$ \cite{Hosur:2015ylk,Maldacena:2015waa}.

The Renyi-2 version entropy is defined by
\begin{equation}
I^{(2)}(A: B: C)=I^{(2)}(A: B)+I^{(2)}(A: C)-I^{(2)}(A: B C)
\end{equation}
where the bipartite mutual information is determined by the Renyi entropy in a way $I^{(2)}(A: C)=$ $S^{(2)}\left(\rho_{A}\right)+S^{(2)}\left(\rho_{C}\right)-S^{(2)}\left(\rho_{A C}\right)$ for $S^{(\alpha)}=\frac{1}{1-\alpha} \log \operatorname{tr}\left(\rho^{\alpha}\right)=\frac{1}{1-\alpha} \log \sum\left(\lambda_{n}\right)^{\alpha}$.

By using the definition of the mutual information from Renyi entropy, we have
\begin{equation}
\begin{aligned}
I_{G H Z}^{(2)}(A: B: C)= & 1-\log _{2}\left[\sum_{n=0}^{\infty} \tanh ^{4 n} \gamma\left(W_{n}(\gamma)-1\right)^{2}\right]-\log _{2}\left[\sum_{n=0}^{\infty} \tanh ^{4 n} \gamma\left(W^{\prime}{ }_{n}(\gamma)-1\right)^{2}\right] \\
& +\log _{2}\left[\sum_{n=0}^{\infty} \tanh ^{4 n} \gamma\left\{\left(W_{n}(\gamma)-2\right)^{2}+1\right\}\right] \\
& +\log _{2}\left[\sum_{n=0}^{\infty} \tanh ^{4 n} \gamma\left\{\left(W^{\prime}(\gamma)-2\right)^{2}+1\right\}\right]\,,
\end{aligned}
\end{equation}
\begin{equation}
\begin{aligned}
I_{W}^{(2)}(A: B: C)= & \log _{2} 9-\log _{2} 5-\log _{2}\left[\sum_{n=0}^{\infty} \tanh ^{4 n} \gamma W_{n}(\gamma)\right]-\log _{2}\left[\sum_{n=0}^{\infty} \tanh ^{4 n} \gamma W^{\prime}{ }_{n}(\gamma)\right] \\
& +2 \log _{2}\left[1+\sum_{n=0}^{\infty} \tanh ^{4 n} \gamma\left(\left(\frac{M_{n}^{+}(\gamma)}{2}\right)^{2}+\left(\frac{M_{n}^{-}(\gamma)}{2}\right)^{2}\right)\right]\,.
\end{aligned}
\end{equation}

In these results, it is evident that the behavior of the Renyi-2 version of mutual information aligns with that of the tripartite mutual information calculated using von Neumann entropy, as illustrated in Figures 6 and 7. It is further confirmed that the Renyi-$\alpha$ version of mutual information converges to the tripartite mutual information with von Neumann entropy as $\alpha$ approaches 1.

Similar to the tripartite mutual information, The Renyi-$\alpha$ mutual information for the W state has a positive value at a certain range of the expansion rate $\gamma$ while the Renyi-$\alpha$ mutual information for GHZ state is negative for all $\gamma$. As discussed in section \ref{section 3.2}, there is a difference between the GHZ state and the W state: the negativity is always zero for the GHZ state, whereas the negativity of the W state can have a non-zero value for some $\gamma$. Based on the partial transpose criterion, the non-zero value of the negativity indicates the existence of the distillable entanglement in the quantum state. Even though the negativity is zero, there might be a bound entanglement or quantum correlation. Using this observation on the negativity, we expect that the existence of the entanglement and quantum correlation might explain the different behavior of the Renyi-$\alpha$ version of mutual information between the GHZ state and W state, as shown in Figures 6 and 7.
\begin{figure}
    \centering
    \includegraphics[width=0.7\textwidth]{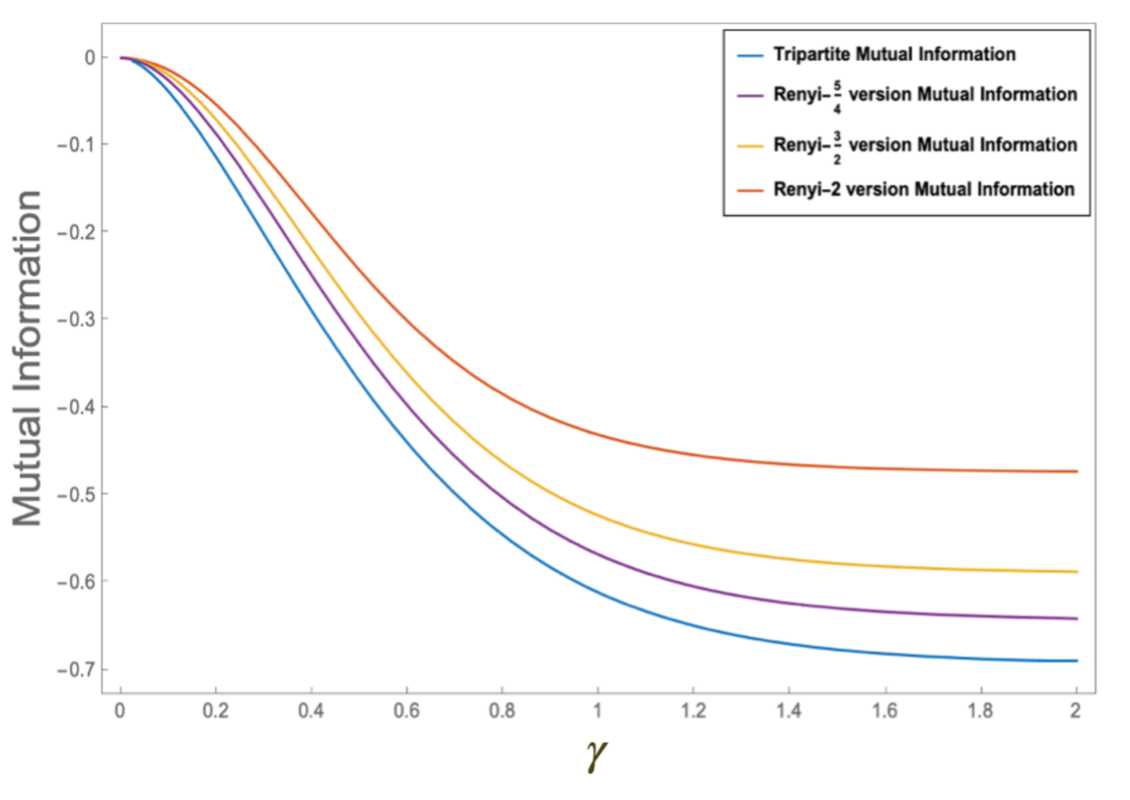}
    \caption{The comparison of the Renyi version of mutual information and the tripartite mutual information with von Neumann entropy (blue line) for the GHZ state reveals in the figure. The figure implies that the I3 scrambling in our paper is consistent with the butterfly effect which can be regarded by the Renyi-2 version mutual information (red line). As shown in the figure, the Renyi version of mutual information converges on the tripartite mutual information in the limit of $\alpha\rightarrow 1$.}
    \label{fig-6}
\end{figure}
\begin{figure}
    \centering
    \includegraphics[width=0.7\textwidth]{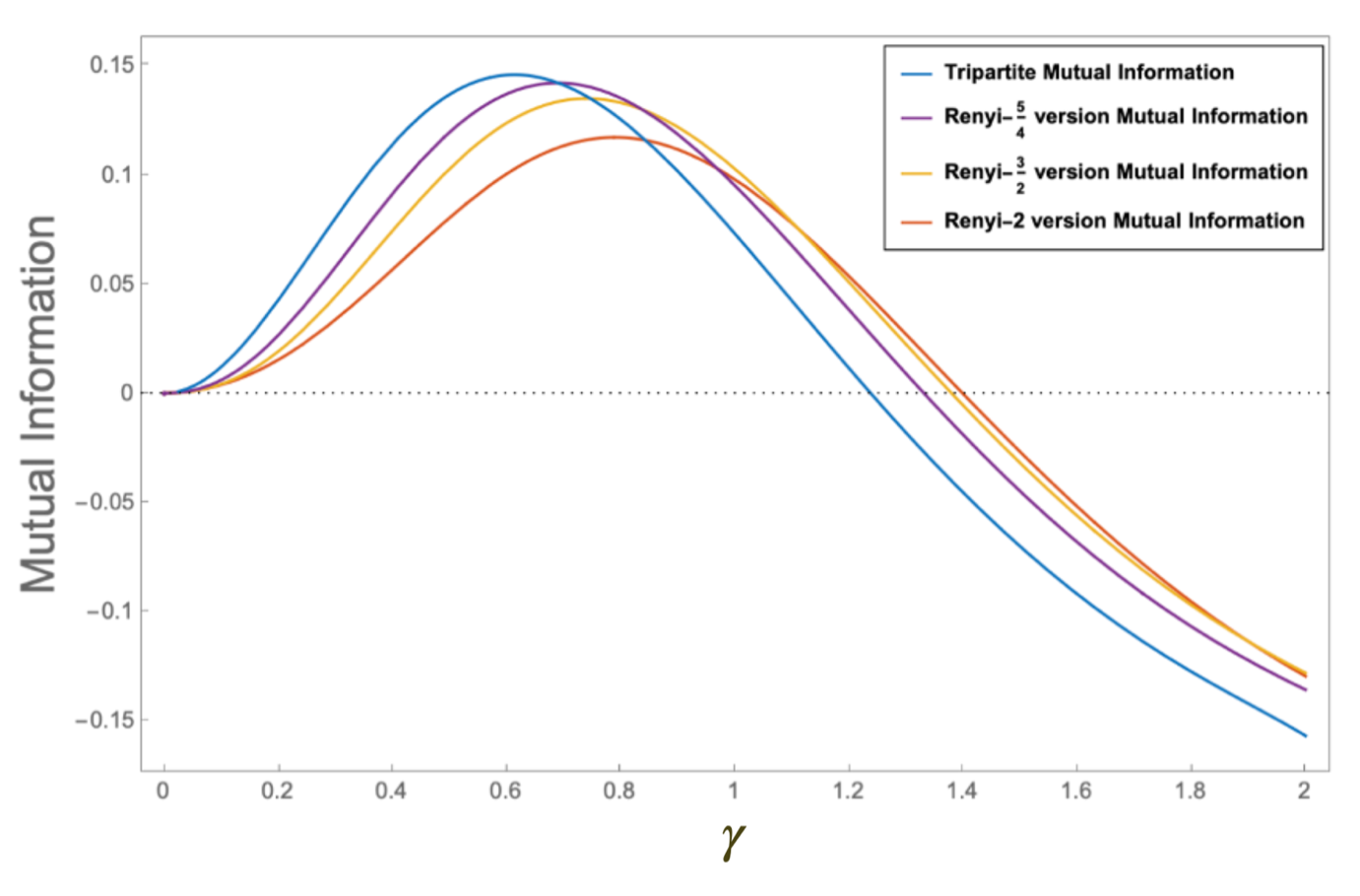}
    \caption{The comparison of the Renyi version of mutual information and the tripartite mutual information with von Neumann entropy (blue line) for the W state reveals in the figure. The figure implies that the I3 scrambling in our paper is consistent with the butterfly effect which can be regarded by the Renyi-2 version mutual information (red line). As shown in the figure, the Renyi version of mutual information converges on the tripartite mutual information in the limit of $\alpha\rightarrow 1$.}
    \label{fig-7}
\end{figure}

We conclude this section by discussing our results. In \cite{Hosur:2015ylk}, it was noted that the tripartite mutual information has a large negative value when there is a significant degree of scrambling in the quantum system. This is consistent with the behavior of (\ref{eq_TMI GHZ}) and (\ref{eq_TMI W}) for large $\gamma$. In our setup, there is a relationship between bipartite mutual information and tripartite mutual information. When the bipartite mutual information has a small magnitude, the tripartite mutual information of the entangled state exhibits a large negative value. This aligns with the understanding that a greater scrambling effect signifies increased difficulty in retrieving information through local measurements. Additionally, it was noted that the tripartite mutual information is less than the Renyi-$\alpha$ version of mutual information for the scrambling system which has negative tripartite mutual information \cite{Hosur:2015ylk}. The aforementioned evidence can be used to strengthen the argument regarding the scrambling effect in de Sitter space \cite{Susskind:1993if}.




\section{Conclusion}\label{section_4}

We study the quantum correlations for tripartite entangled states, focusing on massless scalar particles interacting with de Sitter spacetime. We begin by examining the quantum correlations using the noisy quantum channel. The expansion's effects can be characterized by an operator sum representation, accompanied by the corresponding Kraus operator. This map is demonstrated to be trace-preserving and a completely positive map. Utilizing these two properties, we interpret the quantum effects in de Sitter spacetime as a noisy quantum channel.

We then employ channel-state correspondence to investigate entanglement degradation as a function of universe expansion. This reduction is due to Bob detecting thermal radiation from the de Sitter horizon, while an asymptotically non-vanishing entanglement still remains. For a high expansion rate, the tripartite mutual information has a large negative value. In this regime, we found that a small magnitude of bipartite mutual information is observed when the tripartite mutual information has a large magnitude. This aligns with the idea that scrambling implies difficulty in recovering information from local measurements, leading us to suggest that there is a scrambling effect in de Sitter space. Finding direct evidence of the scrambling effect in the de Sitter space is an interesting topic to investigate, as it could potentially lead to a better understanding of the connections between quantum mechanics, gravity, and cosmology.

Furthermore, we find non-trivial behavior for the tripartite mutual information of the W state at low expansion rates. This suggests that the remaining entanglement in a thermalized W state at a low expansion rate $\gamma$ is connected to the non-trivial behavior of tripartite mutual information. We propose that this behavior is associated with the existence of bound entanglement. It would be interesting for future work to investigate the bound entanglement in this regime.\\

\noindent
{\bf Acknowledgments}\\
We would like to thank Nakwoo Kim, Ki Hyuk Yee, and Hansol Noh for the helpful discussions. This work was supported by Institute for Information \& communications Technology Promotion (IITP) grant funded by the Korean government (MSIP) co-sponsored by Air Force Office of Scientific Research (AFOSR) (Grant No. 2017-0-00266, Gravitational Effects on the Free Space Quantum Key Distribution for Satellite Communication) and the Institute for Information \& Communications Technology Promotion (IITP) (Grant No. IITP 2019-2015-0-00385, ITRC Center for Quantum Communications). PMA and WAM would like to thank the Air Force Office of Scientific Research for their support. WAM's research was supported in part under AFOSR/AOARD grant No. FA2386-17-1-4070 and AFOSR grant No. FA955019-1-0389. . D. A is also supported by AFOSR grant No. F2386-22-1-4052 and AFOSR grant No. F238621-1-0089. Any opinions, findings, conclusions or recommendations expressed in this material are those of the author(s) and do not necessarily reflect the views of AFRL.

\appendix
\section{Review of channel-state correspondence}\label{Appendix_A}
In Appendix \ref{Appendix_A}, we introduce the concept of channel-state correspondence and its application, drawing upon the work of \cite{Hosur:2015ylk}, which demonstrated that properties of a quantum channel can be investigated through calculations performed on the associated quantum state. Utilizing this approach, we will analyze not only the quantum correlation between subsystems within the channel but also other aspects of their interaction.

In quantum information theory, channel-state correspondence (also known as state-channel duality) is a principle that establishes a connection between quantum channels and quantum states. It highlights the equivalence between certain properties of quantum channels and those of quantum states, allowing one to analyze and understand the behavior of quantum channels by examining the associated quantum states. Quantum channels describe the evolution of quantum states under noisy conditions, and they can be represented by completely positive, trace-preserving (CPTP) maps. Quantum states, on the other hand, are described by density matrices. The channel-state correspondence states that there exists a bipartite state, often called a Choi state, associated with every quantum channel. The Choi state contains complete information about the channel and can be used to extract various properties of the channel, such as its capacity, entanglement properties, and other operational characteristics. By exploiting this correspondence, researchers can apply the tools and techniques developed for analyzing quantum states to study the properties of quantum channels. This approach simplifies the analysis of quantum channels and offers valuable insights into their behavior.

The quantum channel is characterized by the information concerning the input state and output state within the Hilbert spaces $H_{in}$ and $H_{out}$, respectively. With complete information on both the input and output states, the quantum channel can be represented by a matrix corresponding to a unitary operation. By employing the channel-state correspondence, one can derive the quantum channel from a state that contains complete information about the unitary operation, as demonstrated in the following general expression.
\begin{equation}\label{eq_1}
    | \Psi \rangle = \sum_j \sqrt{p_j} |\psi_j\rangle_{in}\otimes |\phi_j\rangle_{out}=\sum_{j}(I\otimes U(t))\sqrt{p_j}|\psi_j\rangle_{in}\otimes|\psi_j\rangle_{out}\,,
\end{equation}
where $U(t)|\psi_j\rangle=|\phi_j\rangle$ is the unitary operation from input state to output state, and $\rho_{in}=\sum_j p_j |\psi_j\rangle\langle\psi_j|$ is a density matrix of the input Hilbert space, $\rho_{out}=\sum_j p_j |\phi_j\rangle\langle\phi_j|$ is a density matrix of the output Hilbert space. In the final term of equation (\ref{eq_1}), the state $|\Psi\rangle$ represents an entangled state that incorporates the unitary operation acting on the entangled states themselves. As a result, the state $|\Psi\rangle$ encapsulates the information regarding the unitary operation, along with the input and output states. Owing to this property, we will utilize this state to investigate the characteristics of the unitary operation

One can examine the subsystems within both the input and output systems. The input and output state can be defined as a composition of the subsystems of the channel which give a convenient diagram of the channel. Here, we consider the bipartite input system which is constructed by $H_{in}=H_A\otimes H_B$. The output system which is made from the unitary operation of the input system can also be considered as a bipartite system $H_{out}=H_C \otimes H_D$. Then, we calculate the reduced density matrix of systems $A$ and $C$ by using the partial trace.\footnote{We will write $A \cup B$ by $AB$ for convenience}
\begin{equation}
    \rho_{AC}=-tr_{BD}(\rho_{ABCD})\,.
\end{equation}
In this approach, we calculate the von Neumann entropy among the subsystems within the channel. The reduced density matrix allows us to express the von Neumann entropy of systems $A$ and $C$ in the following form.
\begin{equation}
    S_{AC}=-tr(\rho_{AC}\log{\rho_{AC}})\,.
\end{equation}
Utilizing a comparable calculation method, we determine the von Neumann entropy for each system, denoted as $S(\rho_A)$ and $S(\rho_C)$.

Subsequently, the mutual information can be computed using the following expression:
\begin{equation}
I(A:C)=S(\rho_A)+S(\rho_C)-S(\rho_{AC})\,.
\end{equation}
This implies the presence of a correlation between subsystem $A$ and subsystem $C$. If we choose a bipartite subsystem that exists exclusively within either the input or output system, the von Neumann entropy is equal to the dimension of the complete Hilbert space.
\begin{equation}\label{eq_5}
S\left(\rho_{A B}\right)=S\left(\rho_{C D}\right)=n\,.
\end{equation}
This indicates that the input state is maximally entangled with the output state in the channel-state correspondence \cite{Hosur:2015ylk} when equation (\ref{eq_5}) holds, i.e., if the input state defined in equation (\ref{eq_1}) is of the form $p_j=1/\sqrt{n}$. Equation (\ref{eq_5}) is only true if the input state defined in equation (\ref{eq_1}) is of the form $p_j=1/\sqrt{n}$.  Additionally, this fact suggests that when we select a subsystem exclusively within the input Hilbert space or solely within the output Hilbert space, we can obtain the maximally mixed state. We also note that the channel-state correspondence defined in equation (\ref{eq_1}) only applies to a unitary channel rather than a general quantum channel.

For example, consider a system of 2-qubit system with a swap channel.

\begin{equation}
|i\rangle_{A}|j\rangle_{B} \rightarrow|j\rangle_{C}|i\rangle_{D}\,.
\end{equation}
Here, $|i\rangle_{A}|j\rangle_{B} \in H_{A} \otimes H_{B}=H_{\text {in }}$, and $|j\rangle_{C}|i\rangle_{D} \in H_{C} \otimes H_{D}=H_{\text {out }}$. Then, the state is expressed as
\begin{equation}
|\Psi\rangle=\frac{1}{4} \sum_{i, j=0}^{1}\left(|i\rangle_{A}|j\rangle_{B}\right) \otimes\left(|j\rangle_{C}|i\rangle_{D}\right)\,.
\end{equation}
By using $\rho_{A B C D}=|\Psi\rangle\langle\Psi|$, the reduced density matrices for the systems $AB$ and $A$ yield
\begin{equation}
\begin{gathered}
\rho_{A B}=\frac{1}{4}\left(|0\rangle_{A}\langle 0|+| 1\rangle_{A}\langle 1|\right) \otimes\left(|0\rangle_{B}\langle 0|+| 1\rangle_{B}\langle 1|\right)\,, \\
\rho_{A}=\frac{1}{2}\left(|0\rangle_{A}\langle 0|+| 1\rangle_{A}\langle 1|\right)\,.
\end{gathered}
\end{equation}
The system $CD$ exhibits the same structure of density matrices as the system $AB$. Moreover, systems $B$, $C$, and $D$ share an identical form of density matrices with system $A$. Consequently, the entanglement entropy for each subsystem can be expressed as:
\begin{equation}
S\left(\rho_{A B}\right)=S\left(\rho_{C D}\right)=2, \quad S\left(\rho_{A}\right)=S\left(\rho_{B}\right)=S\left(\rho_{C}\right)=S\left(\rho_{D}\right)=1 .
\end{equation}
In this particular case, $n$ equals 2.

This implies that the system $AB$ and $CD$ are maximally entangled. In this context, the dimension of the Hilbert space for a 2-qubit system is 4, so the entanglement entropy should be $\log _{2} 4=2$. Additionally, subsystem $A$ in the input system is entangled with one of the output subsystems. Given that the dimension of $H_{A}$ is 2, the entanglement entropy in the maximal case should be $\log _{2} 2=1$. The maximal entanglement between $A$ and a subsystem in the output system suggests that system $A$ is maximally mixed. Through this simple example, we verify two facts in state-channel correspondence: (i) the input state and the output state are maximally entangled, and (ii) any subsystem in the input system (or any system in the output system) is maximally mixed.

Interestingly, the scrambling effect can be investigated using tripartite mutual information. Tripartite mutual information is expressed in the following form:

\begin{equation}\label{eq_tripartite mutual information}
I(A: B: C)=I(A: B)+I(A: C)-I(A: B C)
\end{equation}

This illustrates the amount of information obtained about system $A$, through each system $B$ and $C$, rather than the combined system $BC$. This implies that if a local measurement on subsystems $B$ or $C$ provides less information about subsystem $A$, then the tripartite mutual information will have a significantly negative value. In this context, scrambling refers to the challenges associated with retrieving information through local measurements. Specifically, when selecting a subsystem with a dimension less than half of the total Hilbert space, any choice ultimately results in a maximally mixed state. This suggests that the information of some subsystems cannot be decoded through local measurements. Consequently, a large negative value of tripartite mutual information signifies a significant scrambling effect \cite{Hosur:2015ylk}.


\begin{thebibliography}{10}

\bibitem{nielsen2002quantum}
M.~A. Nielsen and I.~L. Chuang, Quantum computation and quantum information, Cambridge Univ. Press, Cambridge (2010), \href{http://dx.doi.org/10.1017/CBO9780511976667}{http://dx.doi.org/10.1017/CBO9780511976667}.

\bibitem{PhysRevA.54.2614}
B.~Schumacher, Sending entanglement through noisy quantum channels, Phys. Rev. A {54} (1996) 2614--2628 \href{http://dx.doi.org/10.1103/PhysRevA.54.2614}{http://dx.doi.org/10.1103/PhysRevA.54.2614}.

\bibitem{Alsing:2003es}
P.~M. Alsing and G.~J. Milburn, Teleportation with a uniformly accelerated partner, Phys. Rev. Lett.  {91} (2003) 180404, \href{http://dx.doi.org/10.1103/PhysRevLett.91.180404}{http://dx.doi.org/10.1103/PhysRevLett.91.180404}.

\bibitem{Fuentes-Schuller:2004iaz}
I.~Fuentes-Schuller and R.~B. Mann, Alice falls into a black hole: Entanglement in non-inertial frames, Phys. Rev. Lett. { 95} (2005) 120404, \href{http://dx.doi.org/10.1103/PhysRevLett.95.120404}{http://dx.doi.org/10.1103/PhysRevLett.95.120404}.

\bibitem{Alsing:2006cj}
P.~M. Alsing, I.~Fuentes-Schuller, R.~B. Mann and T.~E. Tessier, Entanglement of Dirac fields in non-inertial frames, Phys. Rev. A {74} (2006) 032326, \href{http://dx.doi.org/10.1103/PhysRevA.74.032326}{http://dx.doi.org/10.1103/PhysRevA.74.032326}.

\bibitem{DoyeolAhn:2017jld}
D.~Ahn, Unruh effect as a noisy quantum channel, Phys. Rev. A {98} (2018) 022308, \href{http://dx.doi.org/10.1103/PhysRevA.98.022308}{http://dx.doi.org/10.1103/PhysRevA.98.022308}.

\bibitem{DoyeolAhn:2018tyo}
D.~Ahn, Hawking effects as a noisy quantum channel, J. Korean Phys. Soc. {72} (2018) 201--207, \href{http://dx.doi.org/10.3938/jkps.72.201}{http://dx.doi.org/10.3938/jkps.72.201}.

\bibitem{Ball:2005xa}
J.~L. Ball, I.~Fuentes-Schuller and F.~P. Schuller, Entanglement in an expanding spacetime, Phys. Lett. A {359} (2006) 550--554, \href{http://dx.doi.org/10.1016/j.physleta.2006.07.028}{http://dx.doi.org/10.1016/j.physleta.2006.07.028}.

\bibitem{Fuentes:2010dt}
I.~Fuentes, R.~B. Mann, E.~Martin-Martinez and S.~Moradi, {{Entanglement of Dirac fields in an expanding spacetime}}, {{Phys. Rev. D} {82} (2010) 045030}, \href{http://dx.doi.org/10.1103/PhysRevD.82.045030}{http://dx.doi.org/10.1103/PhysRevD.82.045030}.

\bibitem{Nambu:2011ae}
Y.~Nambu and Y.~Ohsumi, {{Classical and Quantum Correlations of Scalar Field in the Inflationary Universe}}, {{Phys. Rev. D} {84} (2011) 044028}, \href{http://dx.doi.org/10.1103/PhysRevD.84.044028}{http://dx.doi.org/10.1103/PhysRevD.84.044028}.

\bibitem{Alsing:2005dno}
P.~M. Alsing, J.~P. Dowling and G.~J. Milburn, {{Ion Trap Simulations of Quantum Fields in an Expanding Universe}}, {{Phys. Rev. Lett.} {94} (2005) 220401}, \href{http://dx.doi.org/10.1103/PhysRevLett.94.220401} {http://dx.doi.org/10.1103/PhysRevLett.94.220401}.

\bibitem{Kanno:2014ifa}
S.~Kanno, {{Impact of quantum entanglement on spectrum of cosmological fluctuations}}, {{JCAP} {07} (2014) 029}, \href{http://dx.doi.org/10.1088/1475-7516/2014/07/029}{http://dx.doi.org/10.1088/1475-7516/2014/07/029}.

\bibitem{Maldacena:2012xp}
J.~Maldacena and G.~L. Pimentel, {{Entanglement entropy in de Sitter space}}, {{JHEP} {02} (2013) 038}, \href{http://dx.doi.org/10.1007/JHEP02(2013)038}{http://dx.doi.org/10.1007/JHEP02(2013)038}.

\bibitem{Choudhury:2016cso}
S.~Choudhury, S.~Panda and R.~Singh, {{Bell violation in the Sky}}, {{Eur. Phys. J. C} {77} (2017) 60}, \href{http://dx.doi.org/10.1140/epjc/s10052-016-4553-3}{http://dx.doi.org/10.1140/epjc/s10052-016-4553-3}.

\bibitem{Choudhury:2016pfr}
S.~Choudhury, S.~Panda and R.~Singh, {{Bell violation in primordial cosmology}}, {{Universe} {3} (2017) 13}, \href{http://dx.doi.org/10.3390/universe3010013}{http://dx.doi.org/10.3390/universe3010013}.

\bibitem{Choudhury:2017bou}
S.~Choudhury and S.~Panda, {{Entangled de Sitter from stringy axionic Bell pair I: an analysis using Bunch\textendash{}Davies vacuum}}, {{Eur. Phys. J. C} {78} (2018) 52}, \href{http://dx.doi.org/10.1140/epjc/s10052-017-5503-4}{http://dx.doi.org/10.1140/epjc/s10052-017-5503-4}.

\bibitem{Choudhury:2017qyl}
S.~Choudhury and S.~Panda, {{Quantum entanglement in de Sitter space from stringy axion: An analysis using $\alpha$ vacua}}, {{Nucl. Phys. B} {943} (2019) 114606}, \href{http://dx.doi.org/10.1016/j.nuclphysb.2019.03.018}{http://dx.doi.org/10.1016/j.nuclphysb.2019.03.018}.

\bibitem{Choudhury:2018ppd}
S.~Choudhury and S.~Panda, {{Cosmological Spectrum of Two-Point Correlation Function from Vacuum Fluctuation of Stringy Axion Field in De Sitter Space: A Study of the Role of Quantum Entanglement}}, {{Universe} {6} (2020) 79}, \href{http://dx.doi.org/10.3390/universe6060079}{http://dx.doi.org/10.3390/universe6060079}.

\bibitem{Bohra2021}
H.~Bohra, S.~Choudhury, P.~Chauhan, P.~Narayan, S.~Panda and A.~Swain, {{Relating the curvature of De Sitter universe to open quantum Lamb shift spectroscopy}}, {{Eur. Phys. J. C} {81} (2021) 196}, \href{http://dx.doi.org/10.1140/epjc/s10052-021-08977-1}{http://dx.doi.org/10.1140/epjc/s10052-021-08977-1}.

\bibitem{Akhtar:2019qdn}
S.~Akhtar, S.~Choudhury, S.~Chowdhury, D.~Goswami, S.~Panda and A.~Swain, {{Open Quantum Entanglement: A study of two atomic system in static patch of de Sitter space}}, {{Eur. Phys. J. C} {80} (2020) 748}, \href{http://dx.doi.org/10.1140/epjc/s10052-020-8302-2}{http://dx.doi.org/10.1140/epjc/s10052-020-8302-2}.

\bibitem{Banerjee:2020ljo}
S.~Banerjee, S.~Choudhury, S.~Chowdhury, R.~N. Das, N.~Gupta, S.~Panda and A.~Swain, {{Indirect detection of Cosmological Constant from interacting open quantum system}}, {{Annals Phys.} {443} (2022) 168941}, \href{http://dx.doi.org/10.1016/j.aop.2022.168941}{http://dx.doi.org/10.1016/j.aop.2022.168941}.

\bibitem{Choudhury:2022mch}
S.~Choudhury, {{Entanglement negativity in de Sitter biverse from Stringy Axionic Bell pair: An analysis using Bunch-Davies vacuum}}, \href{https://doi.org/10.48550/arXiv.2301.05203}{
https://doi.org/10.48550/arXiv.2301.05203}.

\bibitem{Kanno:2014lma}
S.~Kanno, J.~Murugan, J.~P. Shock and J.~Soda, {{Entanglement entropy of $\alpha$-vacua in de Sitter space}}, {{JHEP} {07} (2014) 072}, \href{http://dx.doi.org/10.1007/JHEP07(2014)072}{http://dx.doi.org/10.1007/JHEP07(2014)072}.

\bibitem{Kanno:2014bma}
S.~Kanno, J.~P. Shock and J.~Soda, {{Entanglement negativity in the multiverse}}, {{JCAP} {03} (2015) 015}, \href{http://dx.doi.org/10.1088/1475-7516/2015/03/015}{http://dx.doi.org/10.1088/1475-7516/2015/03/015}.

\bibitem{Kanno:2015ewa}
S.~Kanno, {{A note on initial state entanglement in inflationary cosmology}}, {{EPL} {111} (2015) 60007}, \href{http://dx.doi.org/10.1209/0295-5075/111/60007}{http://dx.doi.org/10.1209/0295-5075/111/60007}.

\bibitem{Kanno:2016gas}
S.~Kanno, J.~P. Shock and J.~Soda, {{Quantum discord in de Sitter space}}, {{Phys. Rev. D} {94} (2016) 125014}, \href{http://dx.doi.org/10.1103/PhysRevD.94.125014}{http://dx.doi.org/10.1103/PhysRevD.94.125014}.

\bibitem{Kanno:2016qcc}
S.~Kanno, M.~Sasaki and T.~Tanaka, {{Vacuum State of the Dirac Field in de Sitter Space and Entanglement Entropy}}, {{JHEP} {03} (2017)
  068}, \href{http://dx.doi.org/10.1007/JHEP03(2017)068}{http://dx.doi.org/10.1007/JHEP03(2017)068}.

\bibitem{Kanno:2017dci}
S.~Kanno and J.~Soda, {{Infinite violation of Bell inequalities in inflation}}, {{Phys. Rev. D} {96} (2017) 083501}, \href{http://dx.doi.org/10.1103/PhysRevD.96.083501}{http://dx.doi.org/10.1103/PhysRevD.96.083501}.

\bibitem{Albrecht:2018prr}
A.~Albrecht, S.~Kanno and M.~Sasaki, {{Quantum entanglement in de Sitter space with a wall, and the decoherence of bubble universes}}, {{Phys. Rev. D} {97} (2018) 083520}, \href{http://dx.doi.org/10.1103/PhysRevD.97.083520}{http://dx.doi.org/10.1103/PhysRevD.97.083520}.

\bibitem{Kanno:2021gpt}
S.~Kanno, J.~Soda and J.~Tokuda, {{Indirect detection of gravitons through quantum entanglement}}, {{Phys. Rev. D} {104} (2021) 083516}, \href{http://dx.doi.org/10.1103/PhysRevD.104.083516}{http://dx.doi.org/10.1103/PhysRevD.104.083516}.

\bibitem{Kanno:2022kve}
S.~Kanno, A.~Mukuno, J.~Soda and K.~Ueda, {{Impact of quantum entanglement induced by magnetic fields on primordial gravitational waves}}, {{Phys. Rev. D} {107} (2023) 063503}, \href{http://dx.doi.org/10.1103/PhysRevD.107.063503}{http://dx.doi.org/10.1103/PhysRevD.107.063503}.

\bibitem{Torres-Arenas:2018vei}
A.~J. Torres-Arenas, Q.~Dong, G.-H. Sun and S.-H. Dong, {{Entanglement measures of W-state in noninertial frames}}, {{Phys. Lett. B} {789} (2019) 93}, \href{http://dx.doi.org/10.1016/j.physletb.2018.12.010}{http://dx.doi.org/10.1016/j.physletb.2018.12.010}.

\bibitem{Torres-Arenas_2019}
A.~J. Torres-Arenas, E.~O. López-Zúñiga, J.~A. Saldaña-Herrera, Q.~Dong, G.-H. Sun and S.-H. Dong, {Tetrapartite entanglement measures of W-class in noninertial frames}, {{Chinese Physics B} {28} (2019) 070301}, \href{http://dx.doi.org/10.1088/1674-1056/28/7/070301}{http://dx.doi.org/10.1088/1674-1056/28/7/070301}.

\bibitem{Quantum_2019}
W.-C. Qiang, Q.~Dong, M.~A.~M. Sanchez, G.-H. Sun and S.-H. Dong, {Entanglement property of the Werner state in accelerated frames}, {{Quantum Information Processing} {18} (2019) 314}, \href{http://dx.doi.org/10.1007/s11128-019-2421-4}{http://dx.doi.org/10.1007/s11128-019-2421-4}.

\bibitem{Dong_2019}
Q.~Dong, M.~A.~M. Sanchez, G.-H. Sun, M.~Toutounji and S.-H. Dong, {Tripartite entanglement measures of generalized GHZ state in uniform acceleration}, {{Chinese Physics Letters} {36} (2019) 100301}, \href{http://dx.doi.org/10.1088/0256-307X/36/10/100301}{http://dx.doi.org/10.1088/0256-307X/36/10/100301}.

\bibitem{Dong:2019iqt}
Q.~Dong, A.~J. Torres-Arenas, G.-H. Sun and S.-H. Dong, {{Tetrapartite entanglement features of W-Class state in uniform acceleration}}, {{Front. Phys.} {15} (2020) 11602}, \href{http://dx.doi.org/10.1007/s11467-019-0940-1}{http://dx.doi.org/10.1007/s11467-019-0940-1}.

\bibitem{Dong:2022vfw}
Q.~Dong, R.~S. Carrillo, G.-H. Sun and S.-H. Dong, {{Tetrapartite entanglement measures of generalized GHZ state in the noninertial frames}}, {{Chin. Phys. B} {31} (2022) 030303}, \href{http://dx.doi.org/10.1088/1674-1056/ac2299}{http://dx.doi.org/10.1088/1674-1056/ac2299}.

\bibitem{Hosur:2015ylk}
P.~Hosur, X.-L. Qi, D.~A. Roberts and B.~Yoshida, {{Chaos in quantum channels}}, {{JHEP} {02} (2016) 004}, \href{http://dx.doi.org/10.1007/JHEP02(2016)004}{http://dx.doi.org/10.1007/JHEP02(2016)004}.

\bibitem{Gibbons:1977mu}
G.~W. Gibbons and S.~W. Hawking, {{Cosmological Event Horizons, Thermodynamics, and Particle Creation}}, {{Phys. Rev. D} {15} (1977) 2738--2751}, \href{http://dx.doi.org/10.1103/PhysRevD.15.2738}{http://dx.doi.org/10.1103/PhysRevD.15.2738}.

\bibitem{Lohiya:1978}
D.~Lohiya and N.~Panchapakesan, {Massless scalar field in a de sitter universe and its thermal flux}, {{Journal of Physics A: Mathematical and General} {11} (1978) 1963}, \href{http://dx.doi.org/10.1088/0305-4470/11/10/014}{http://dx.doi.org/10.1088/0305-4470/11/10/014}.

\bibitem{Birrell:1982ix}
N.~D. Birrell and P.~C.~W. Davies, {{Quantum Fields in Curved Space}}, Cambridge Univ. Press, Cambridge (1982), \href{http://dx.doi.org/10.1017/CBO9780511622632}{http://dx.doi.org/10.1017/CBO9780511622632}.

\bibitem{Unruh:1976db}
W.~G. Unruh, {{Notes on black hole evaporation}}, {{Phys. Rev. D} {14} (1976) 870}, \href{http://dx.doi.org/10.1103/PhysRevD.14.870}{http://dx.doi.org/10.1103/PhysRevD.14.870}.

\bibitem{Dur:2000zz}
W.~Dur, G.~Vidal and J.~I. Cirac, {{Three qubits can be entangled in two inequivalent ways}}, {{Phys. Rev. A} {62} (2000) 062314}, \href{http://dx.doi.org/10.1103/PhysRevA.62.062314}{http://dx.doi.org/10.1103/PhysRevA.62.062314}.

\bibitem{Vidal:2002zz}
G.~Vidal and R.~F. Werner, {{Computable measure of entanglement}}, {{Phys. Rev. A} {65} (2002) 032314}, \href{http://dx.doi.org/10.1103/PhysRevA.65.032314}{http://dx.doi.org/10.1103/PhysRevA.65.032314}.

\bibitem{Horodecki:2009zz}
R.~Horodecki, P.~Horodecki, M.~Horodecki and K.~Horodecki, {{Quantum entanglement}}, {{Rev. Mod. Phys.} { 81} (2009) 865--942}, \href{http://dx.doi.org/10.1103/RevModPhys.81.865}{http://dx.doi.org/10.1103/RevModPhys.81.865}.

\bibitem{Hill:1997pfa}
S.~Hill and W.~K. Wootters, {{Entanglement of a pair of quantum bits}}, {{Phys. Rev. Lett.} {78} (1997) 5022--5025}, \href{http://dx.doi.org/10.1103/PhysRevLett.78.5022}{http://dx.doi.org/10.1103/PhysRevLett.78.5022}.

\bibitem{Rungta_2001}
P.~Rungta, V.~Bu{\v{z} }ek, C.~M. Caves, M.~Hillery and G.~J. Milburn, {Universal state inversion and concurrence in arbitrary dimensions}, {{Physical Review A} {64} (2001)}, \href{http://dx.doi.org/10.1103/physreva.64.042315}{http://dx.doi.org/10.1103/physreva.64.042315}.

\bibitem{Maldacena:2015waa}
J.~Maldacena, S.~H. Shenker and D.~Stanford, {{A bound on chaos}}, {{JHEP} {08} (2016) 106}, \href{http://dx.doi.org/10.1007/JHEP08(2016)106}{http://dx.doi.org/10.1007/JHEP08(2016)106}.

\bibitem{Susskind:1993if}
L.~Susskind, L.~Thorlacius and J.~Uglum, {{The Stretched horizon and black hole complementarity}}, {{Phys. Rev. D} {48} (1993) 3743--3761}, \href{http://dx.doi.org/10.1103/PhysRevD.48.3743}{http://dx.doi.org/10.1103/PhysRevD.48.3743}.

\end{thebibliography}
\end{document}